\documentclass[english,preprint, aps]{revtex4-1}
\usepackage[LGR,T1]{fontenc}
\usepackage[latin9]{inputenc}
\usepackage{array}
\usepackage{booktabs}
\usepackage{units}
\usepackage{textcomp}
\usepackage{multirow}
\usepackage{amsmath}
\usepackage{amssymb}
\usepackage{graphicx}
\usepackage{setspace}

\makeatletter

\DeclareRobustCommand{\greektext}{%
  \fontencoding{LGR}\selectfont\def\encodingdefault{LGR}}
\DeclareRobustCommand{\textgreek}[1]{\leavevmode{\greektext #1}}
\ProvideTextCommand{\~}{LGR}[1]{\char126#1}

\providecommand{\tabularnewline}{\\}

\@ifundefined{showcaptionsetup}{}{%
 \PassOptionsToPackage{caption=false}{subfig}}
\usepackage{subfig}
\makeatother

\usepackage{babel}
\begin{document}

\title{Projections for Neutral Di-Boson and Di-Higgs Interactions at FCC-he
Collider}

\author{S. Kuday}
\email{sinankuday@aydin.edu.tr}

\affiliation{Istanbul Ayd\i n University, Application and Research Center For
Advanced Studies, 34295, Istanbul, Turkey}

\author{H. Sayg\i n}
\email{hasansaygin@aydin.edu.tr}

\affiliation{Istanbul Ayd\i n University, Application and Research Center For
Advanced Studies, 34295, Istanbul, Turkey}

\author{\.{I}. Ho\c{s}}
\email{ilknurhos@aydin.edu.tr}

\affiliation{Istanbul Ayd\i n University, Application and Research Center For
Advanced Studies, 34295, Istanbul, Turkey}

\author{F. Çetin}
\email{fusuncetin@aydin.edu.tr}

\affiliation{Istanbul Ayd\i n University, Application and Research Center For
Advanced Studies, 34295, Istanbul, Turkey}
\begin{abstract}
As a high energy e-p collider, FCC-he, has been recently proposed
with sufficient energy options to investigate Higgs couplings. To
analyse the sensitivity on Higgs boson couplings, we focus spesifically
on the CP-even and CP-odd Wilson coefficients with $hhZZ\:$and $hh\gamma\gamma\:$four-point
interactions of Higgs boson with Effective Lagrangian Model through
the process $e^{-}p\to hhje^{-}$ . We simulate the related processes
in FCC-he, with 60 GeV and 120 GeV $e^{-}$ beams and 50 TeV proton
beam collisions. We present the exclusion limits on these couplings
both for 68\% and 95\% C.L. in terms of integrated luminosities.
\end{abstract}

\maketitle

\section{INTRODUCTION}

The discovery of the Higgs boson \citep{Higgs,key-1} and the consistency
of Higgs measurements by ATLAS and CMS \citep{ATLAS,CMS} brought
up all available Higgs production and decay channels to an utmost
importance level. Of these channels, arguably the most important ones
are the Higgs-self coupling ($\lambda$) and the anomalous couplings,
since it will show a direct evidence of the Electroweak Symmetry Breaking
(EWSB) mechanism \citep{EWSBMechanism} which is expected to work
as predicted by Standard Model (SM). 

Over the years, extensive studies have shown that it is quite challenging
to observe the Yukawa couplings of Higgs boson to other fermions even
with the correction algorithms at the LHC through gluon-fusion process
due to the enormous SM background \citep{SMbackground}. Although
Vector Boson Fusion (VBF) processes are accesible at the LHC \citep{key-3},
there are studies suggest that it is more feasible to accomplish this
task using linear colliders \citep{key-4} or through ep-collisions
\citep{key-5}. Consequently, searching for Higgs decays at future
colliders became relatively important just because they bring unique
opportunities to fully cover SM scalar sector \citep{key-14}. To
study anomalous couplings, di-Higgs boson production through charged
current (CC) mechanisms are well studied in Large Hadron Electron
Collider (LHeC) and Future Circular Collider (FCC-he) \citep{key-15}
expressing that neutral current (NC) mechanisms have a potential to
enhance the overall Higgs boson signal efficiency. In addition, for
the completeness of the studies based on the Higgs Effective Lagrangian
Model \citep{key-16}, it is quite promising to study on di-higgs
productions via four-point interaction vertices since it contains
aspects for both new physics and SM Higgs studies. However a complete
understanding of Higgs sector can open doors to new physics and particles.
Likewise, it has recently been reported that if additional scalar
bosons exist, they can be interpreted in the effective theory approach
leaving signatures in the final states with a pair of invisible $\chi$
particles that are proposed to be the dark matter candidate \citep{AddScalars}.
According to this approach, newly proposed heavy Higgs boson eventually
decays to a Higgs boson and a pair of $\chi$, causing a distortion
in the $p_{T}$ distribution that are compatible with the observations
at LHC \citep{AddSca2}.

Here, it is considered the electron - proton collision variant of
the FCC-he that is proposed to build on the same site with LHC, as
the future extension of the LHeC. In FCC-he, construction of an Energy
Recovery Linac is proposed to deliver electrons with energies ranging
from $E_{e}=60\:GeV$ to $E_{e}=120\:GeV$, while a proton beam is
provided by a 100 km circular beam pipe and has an maximum proton
energy of $E_{p}=50\:TeV$. 

The main idea in this letter is that by analysing neutral four-point
interactions in FCC-he, one can get rid of a part of the SM background
and get a better detection efficiency by electron tracks involved
in the final state which can be reconstructed efficiently. We studied
Higgs boson couplings at neutral four-point interaction vertices through
Wilson coefficients within the Higgs Effective Lagrangian Model. The
outline of our paper has been prepared as in the following: In section
2, we basically reveal the related Lagragian terms and their phenomenological
interpretations as well as the assumptions of our case. In section
3 and 4, we discuss event productions for signal and background processes
respectively. In section 5, we explain applied event selection criterias
and statistical analysis of data that we obtained from simulation
tools. Finally in section 5, we present our results and exclusion
limits for obtaining related coefficients in FCC-he collisions. 

\section{HIGGS EFFECTIVE LAGRANGIAN (HEL) MODEL}

Since the details of the Higgs sector is not trivial, an effective
field theory (EFT) that covers all related interactions at a given
scale, but not the others that play a role at significantly different
scales, might be a good approach. In EFT models, particularly interactions
at much higher energies than the energy scale of interest are ignored.
So that the underlying physics event at energies below the new physics
scale can be described precisely. In this letter, we studied on the
exclusive Higgs Effective Lagrangian (HEL) Model, that is valid above
a $\varLambda$ scale around TeV order, makes possible to include
dimension-six operators with free parameters, namely, Higgs self-couplings,
Yukawa couplings and Wilson coefficients. In this approach, the complete
Lagrangian is handled by SM Lagrangian and supplemented higher dimensional
operators which are assumed to appear at energies larger than the
effective scale. $\mathcal{L}$, the most general gauge-invariant
total Lagrangian, can be expressed as in the followings with Wilson
coefficients $\bar{c_{i}}$ and independent operators $\mathcal{O_{\mathit{i}}}$
of dimension less than or equal to six.

\begin{equation}
\mathcal{L=\mathcal{L_{SM}}}+\sum_{i}\bar{c}_{i}\mathcal{O_{\mathrm{\mathit{i}}}}=\mathcal{\mathcal{L_{SM}}}+\mathcal{L_{SILH}}+\mathcal{L_{CPV}}+...\label{eq:1}
\end{equation}

After EWSB, the Higgs sector can be expressed as;

\begin{equation}
\mathcal{L_{\mathit{Higgs}}=\mathcal{L^{\mathit{(3)}}\mathit{+\mathcal{L^{\mathit{(4)}}\mathcal{\mathit{+}L^{\mathit{(5)}}\mathit{+\mathcal{L^{\mathit{(6)}}}}}}}}}\label{eq:2}
\end{equation}

where numbers in superscript denotes the set of interactions of a
Higgs boson with a vector boson pair. Related Lagrangians can specifically
be rewritten as follows for the mass basis.

\begin{singlespace}
\begin{eqnarray}
\mathcal{L_{\mathit{hhh}}^{\mathit{(3)}}=}-\frac{m_{H}^{2}}{2v}g_{hhh}^{(1)}h^{3}+\frac{1}{2}g_{hhh}^{(2)}h\partial_{\mu}h\partial^{\mu}h\label{eq:3}
\end{eqnarray}

\begin{equation}
\mathcal{L}{}_{\mathit{hzz}}^{\mathit{(3)}}=-\frac{1}{4}g_{hzz}^{(1)}Z_{\mu\nu}Z^{\mu\nu}h-g_{hzz}^{(2)}Z_{\nu}\partial_{\mu}Z^{\mu\nu}h+\frac{1}{2}g_{hzz}^{(3)}Z_{\mu}Z^{\mu}h-\frac{1}{4}\tilde{g}_{hzz}Z_{\mu\nu}\tilde{Z}^{\mu\nu}h
\end{equation}

\begin{equation}
\mathcal{L}{}_{\mathit{h\gamma\gamma}}^{\mathit{(3)}}=-\frac{1}{4}g_{h\gamma\gamma}F_{\mu\nu}F^{\mu\nu}h-\frac{1}{4}\tilde{g}_{h\gamma\gamma}F_{\mu\nu}\tilde{F}^{\mu\nu}h
\end{equation}

\begin{equation}
\mathcal{L}{}_{\mathit{hhzz}}^{\mathit{(4)}}=-\frac{1}{8}g_{hhzz}^{(1)}Z_{\mu\nu}Z^{\mu\nu}h^{2}-\frac{1}{2}g_{hhzz}^{(2)}Z_{\nu}\partial_{\mu}Z^{\mu\nu}h^{2}+\frac{1}{4}g_{hhzz}^{(3)}Z_{\mu}Z^{\mu}h^{2}-\frac{1}{8}\tilde{g}_{hhzz}Z_{\mu\nu}\tilde{Z}^{\mu\nu}h^{2}
\end{equation}

\begin{equation}
\mathcal{L}{}_{\mathit{hh\gamma\gamma}}^{\mathit{(4)}}=-\frac{1}{8}g_{hh\gamma\gamma}F_{\mu\nu}F^{\mu\nu}h^{2}-\frac{1}{8}\tilde{g}_{hh\gamma\gamma}F_{\mu\nu}\tilde{F}^{\mu\nu}h^{2}
\end{equation}

\end{singlespace}

Here, tilde operator denotes the CP-violating terms, while all other
non-tilde terms are CP-conserving. One can also consider other neutral
four-point interactions such as di-higgs and di-gluon or quartic-self-interaction
of Higgs. But these processes are shown to give no events at FCC-he
collider. Therefore we can describe the general Lagrangian that we
are working on as $\mathcal{L=\mathcal{L_{SM}}}+\mathcal{L_{\mathit{hhh}}^{\mathit{(3)}}}+\mathcal{L}{}_{\mathit{hzz}}^{\mathit{(3)}}+\mathcal{L}{}_{\mathit{h\gamma\gamma}}^{\mathit{(3)}}+\mathcal{L}{}_{\mathit{hhzz}}^{\mathit{(4)}}+\mathcal{L}{}_{\mathit{hh\gamma\gamma}}^{\mathit{(4)}}$
. Several different representations of couplings in Eq. (3-7) are
available via FCNC notation \citep{KappaNotation}. In principle,
we concentrate on gauge basis representations of couplings with Wilson
coefficients as in Table 1 - 2 and take the same notation as explicitly
described in \citep{key-6}. From Table 2, one can see that $g_{hh\gamma\gamma}$
($\tilde{g}_{hh\gamma\gamma}$) strictly corresponds to terms with
only $\bar{c_{\gamma}}$ ($\tilde{c}_{\gamma}$) coefficient, while
$g_{hhzz}$($\tilde{g}_{hhzz}$) indirectly corresponds to terms with
coefficients $\bar{c}_{HB},\bar{c}_{HW},\bar{c}_{\gamma},\bar{c}_{W}$
($\tilde{c}_{HB},\tilde{c}_{HW},\tilde{c_{\gamma}},\tilde{c_{W}}$)
for the first two orders, respectively. And for the third order of
$g_{hhzz}$ , it is seen an explicit dependence to $\bar{c}_{T},\bar{c}_{H},\bar{c}_{\gamma}$.
To scan over these parameters, we explain our strategy in the next
section with the case-spesific assumptions. To understand physical
analysis of EFT explicitly, one must build SILH (Strongly-Interacting
Light Higgs) Lagrangian in terms of indepent operators as shown in
Eq. (1) and described in Ref \citep{key-Grojean}. One can then discuss
the relative effect of the various operators on physical observables
through Wilson coefficients. However, SILH Lagrangian includes only
CP-conserving operators multiplied with Higgs related fields. For
completeness, one should also add a CP-violating Lagrangian as in
Eq. (1) which has the same interactions with SILH Lagrangian but rewritten
with CP-violating coefficients ($\tilde{c}_{HB},\tilde{c}_{HW},\tilde{c_{\gamma}},\tilde{c_{W}}$)
and operators. One of the naive ways of estimating these coefficient
values has been made by power counting after expanding the effective
Lagrangian in the number of fields and derivatives at tree level.
According to power counting for the related terms that we are interested
in, one can estimates;
\begin{equation}
\bar{c}_{6},\:\bar{c}_{H},\:\bar{c_{T}}\:\sim O\left(\frac{v^{2}}{f^{2}}\right)\qquad\bar{c}_{W},\:\bar{c}_{B}\:\sim O\left(\frac{m_{W}^{2}}{M^{2}}\right)\qquad\bar{c}_{HB},\:\bar{c}_{HW},\:\bar{c}_{\gamma},\:\bar{c}_{g}\sim O\left(\frac{m_{W}^{2}}{f^{2}}\right)
\end{equation}

\begin{table}
\caption{Corresponding couplings of a Higgs boson and a pair of neutral bosons
in the mass and gauge basis for Eq. (3) as in Ref \citep{key-16}. }

\begin{tabular}{|c|c|}
\hline 
Mass Basis & Gauge Basis\tabularnewline
\hline 
$g_{hhh}^{(1)}$ & $1+\frac{7}{8}\bar{c}_{6}-\frac{1}{2}\bar{c}_{H}$\tabularnewline
\hline 
$g_{hhh}^{(2)}$ & $\frac{g}{m_{W}}\bar{c}_{H}$\tabularnewline
\hline 
$g_{h\gamma\gamma}$ & $a_{H}-\frac{8g\bar{c}_{\gamma}s_{W}^{2}}{m_{W}}$\tabularnewline
\hline 
$\tilde{g}_{h\gamma\gamma}$ & $-\frac{8g\tilde{c}_{\gamma}s_{W}^{2}}{m_{W}}$\tabularnewline
\hline 
$g_{hzz}^{(1)}$ & $\frac{2g}{c_{W}^{2}m_{W}}[\bar{c}_{HB}s_{W}^{2}-4\bar{c}_{\gamma}s_{W}^{4}+c_{W}^{2}\bar{c}_{HW}]$\tabularnewline
\hline 
$g_{hzz}^{(2)}$ & $\frac{g}{c_{W}^{2}m_{W}}[(\bar{c}_{HW}+\bar{c}_{W})c_{W}^{2}+(\bar{c}_{B}+\bar{c}_{HB})s_{W}^{2}]$\tabularnewline
\hline 
$g_{hzz}^{(3)}$ & $\frac{gm_{W}}{c_{W}^{2}}[1-\frac{1}{2}\bar{c}_{H}-2\bar{c}_{T}+8\bar{c}_{\gamma}\frac{s_{W}^{4}}{c_{W}^{2}}]$\tabularnewline
\hline 
$\tilde{g}_{hzz}$ & $\frac{2g}{c_{W}^{2}m_{W}}[\tilde{c}_{HB}s_{W}^{2}-4\tilde{c}_{\gamma}s_{W}^{4}+c_{W}^{2}\tilde{c}_{HW}]$\tabularnewline
\hline 
\end{tabular}
\end{table}

\begin{table}
\caption{Corresponding couplings of Higgs and neutral boson pairs in the mass
and gauge basis for Eq. (3) as in Ref \citep{key-16}. }

\begin{tabular}{|c|c|c|c|c|}
\hline 
Mass Basis  & $g_{hh\gamma\gamma}$ & $\tilde{g}_{hh\gamma\gamma}$, $\tilde{g}_{hhzz}$ & $g_{hhzz}^{(1)},g_{hhzz}^{(2)}$ & $g_{hhzz}^{(3)}$\tabularnewline
\hline 
\hline 
Gauge Basis & $-\frac{4\bar{c}_{\gamma}g^{2}s_{W}^{2}}{m_{W}^{2}}$ & $\frac{g}{2m_{W}}\left\{ \tilde{g}_{h\gamma\gamma},\tilde{g}_{hzz}\right\} $ & $\frac{g}{2m_{W}}\{g_{hzz}^{(1)},g_{hzz}^{(2)}\}$ & $\frac{g^{2}}{2c_{W}^{2}}[1-6\bar{c}_{T}-\bar{c}_{H}+8\bar{c}_{\gamma}\frac{s_{W}^{4}}{c_{W}^{2}}]$\tabularnewline
\hline 
\end{tabular}
\end{table}

where $v$ is vacuum expectation value, f denotes the coupling strength
of the Higgs boson to New Physics states and M is the overall mass
scale. If one defines the new physics coupling as $g_{NP}$, f can
explicitly be written as $\nicefrac{g_{NP}}{M}$. Above the tree level,
related coefficient will be shifted upper values slightly getting
contributions from extra terms depent on the mass scale, M. Although
we shall comment on these effects in the conclusion, it is should
be noted that the further evaluations and analysis on the topic are
beyond the scope of this letter. 

\section{SIGNAL PRODUCTION}

For signal production, we have used the implementation of a Higgs
Effective Field Theory in MadGraph5 Model \citep{UFOModel} with FeynRules
\citep{FeynRules} that is avaliable including full lagrangian terms
and a set of independent dimension-six operators. As shown in Fig.1,
we produced events of $e^{-}p\to hhje^{-}$ processes using HEL model
taking into account effective vertices and keeping e-p collider set
up at $\sqrt{s}\approx3.5\:TeV$ and $\sqrt{s}\approx5\:TeV$ which
are the two main options of FCC-he. 
\begin{figure}
\includegraphics{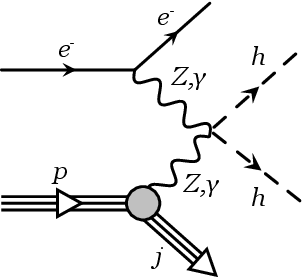} \includegraphics{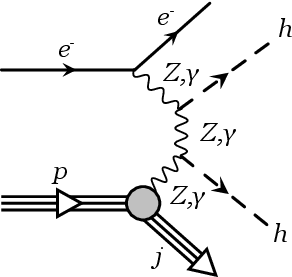}
\includegraphics{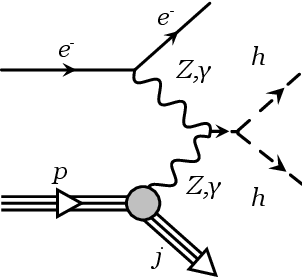}

\caption{Feynman Diagrams for the signal processes}
\end{figure}

\begin{figure}
\subfloat[]{\includegraphics[bb=0bp 0bp 345bp 244bp]{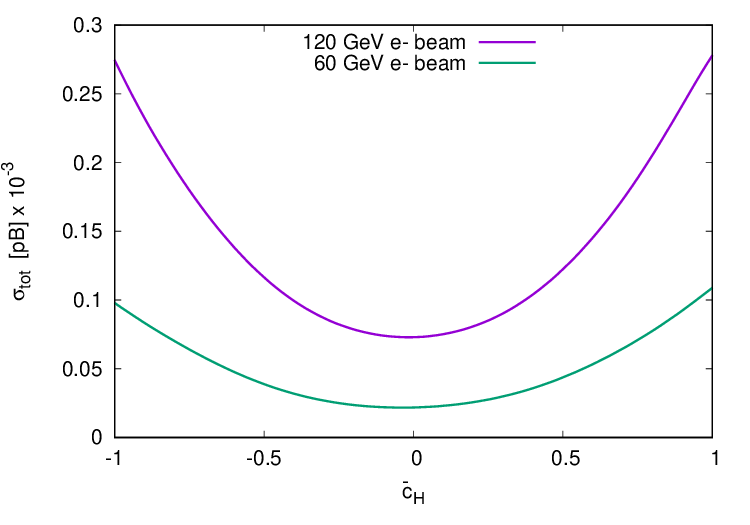}

}\hfill{}\subfloat[]{\includegraphics[bb=0bp 0bp 345bp 244bp]{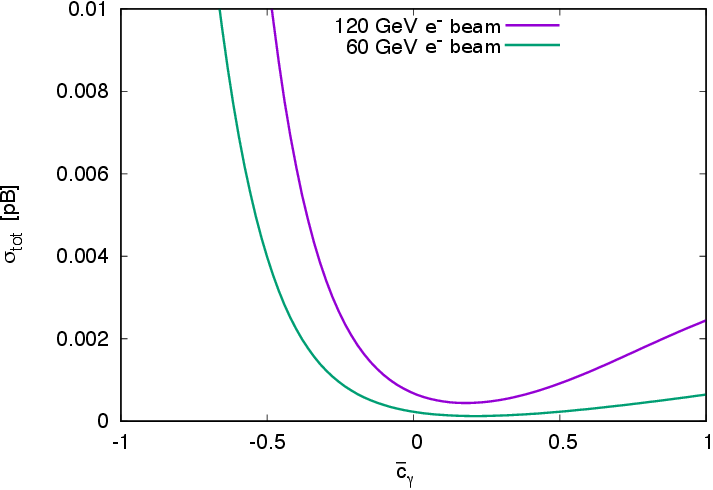} 

}

\caption{NC signal cross section values over scanned (a) $\bar{c}_{H}$(while
$\bar{c}_{\gamma}=0.1$) and (b) $\bar{c}_{\gamma}$(while $\bar{c}_{H}=0.1$)
coefficient for $hhZZ$ vertex coupling through $e^{-}p\to hhje^{-}$
process. }
\end{figure}

\begin{figure}
\includegraphics{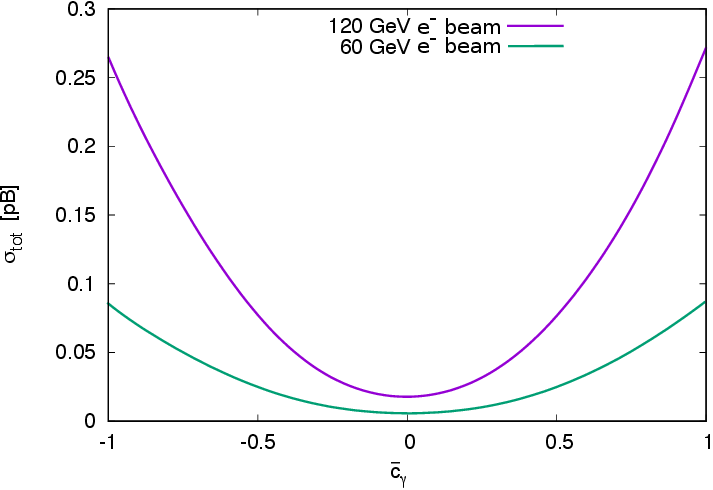}

\caption{PF signal cross section values over scanned $\bar{c}_{\gamma}$ coefficient
for $hh\gamma\gamma$ vertex coupling through $e^{-}p\to hhje^{-}$
process. }
\label{Fig.1}
\end{figure}

Here, we are searching for both $Z$ bosons and $\gamma$ as mediators
that together forms the NC processes. One can name each subprocesses
as Vector Boson Fusion (VBF) with Z boson and photo fusion (PF) with
$\gamma$ mediators, respectively. Due to gauge invariant structure
of HEL model, one can not actually separate event productions for
Z boson and $\gamma$ mediators. However, it is possible to minimize
the contribution of one of the subprocesses setting related Wilson
coefficients as below.  The corresponding couplings of PF process
are well known with already studied in letters Ref \citep{key-PF2,key-PF}
for different colliders. Due to the suppression of the triple higgs
self coupling, one can see that the first two diagrams dominate the
cross section depending on the effective vertices that are defined
by the HEL model. At this point, together with the previous constraints
in the literature, we considered the following restrictions to related
Wilson coefficients during the signal production: 

(1) $\bar{c}_{B}$ and $\bar{c}_{W}$ are suppressed, since they should
be order of $\nicefrac{m_{W}^{2}}{M^{2}}$ where M is the typical
mass scale of the new physics sector. \citep{key-16}

(2) $\bar{c}_{6}$ is also supressed for our case, since the corresponding
production cross section gives no events at FCC-he. 

(3) Constraints from the electroweak precision parameters suggest
that $\bar{c}_{T},\bar{c}_{W},\bar{c}_{B}$ should be order of $10^{-3}$
according to \citep{key-7}. 

(4) $\bar{c}_{W}$ has constrained between {[}\textminus 1.71, 0.42{]}
according to Ref \citep{key-16} extracted from ATLAS results \citep{key-8-1}.

(5) $\bar{c}_{HW}$ and $\bar{c}_{HB}$ that are expected to be order
of $10^{-3}$ tend to cancel each other at the Z-pole. 

We investigated $e^{-}p\to hhje^{-}$ process both for $Z$ boson
and $\gamma$ as mediators for the electron polarization: 0 and -0.8.
It is considered that the main decay channel of higgs as $h\rightarrow b\bar{b}$
and looked for 4b-jets + Singlejet + lepton in the final state. Note
that for PF signal production, we accept $\bar{c}_{T},\bar{c}_{W},\bar{c}_{B}=10^{-3}$
while all other Wilson coefficients are set to zero except scan parameter
$\bar{c_{\gamma}}$. Similarly, for VBF signal production, we accept
$\bar{c}_{T},\bar{c}_{W},\bar{c}_{B}=10^{-3}$ setting all other Wilson
coefficients to zero except scan parameter $\bar{c_{H}}$. Thus, two
signals, VBF and PF are simulated for negative and positive values
of coefficients $\bar{c_{H}}$ and $\bar{c_{\gamma}}$, respectively.
Event data has hadronized by Pythia-PGS \citep{Pythia-Pgs} and detector
level simulation performed by Delphes (version 3.4.1) \citep{Delphes}
that are the packages placed in the MadGraph5\_aMC@NLO \citep{MG5}
framework 2.5.2 release. Recently announced Delphes baseline detector
definitions for FCC-hh have been used to handle the events data by
simulation. For detector definitions, particle propagator defined
with 1.5 m radius, 5 m half length magnetic field coverage and 4 T
z-magnetic field. We assume that the pile-up effects are negligible
at both energy and luminosity options of FCC-he collider. Jets are
clustered with anti-kT clustering algorithm \citep{antiKT} with a
size parameter of $\Delta R=0.5$ by using FastJet package \citep{FastJet}.

From Fig.2 and 3, one can see that a cross section scan over $\bar{c}_{H},\:\bar{c}_{\gamma}$
parameters for the processes where $hhZZ$ and $hh\gamma\gamma$ vertices
involved. It is trivial that PF signal has higher cross sections if
the Wilson coefficient is around $\bar{c_{\gamma}}\eqsim1$. On the
other hand, $hhZZ$ vertex has an asymmetric large sensitivity to
$\bar{c}_{\gamma}$ coefficient as shown in Fig.2b. In Table 3, we
present the event counts for signal and background processes where
both VBF and PF signals are independently produced. 

\section{BACKGROUND PRODUCTION}

Although it is highly supressed in the phase space, one can produce
events for $e^{-}p\rightarrow2(b\bar{b})je^{-}$ process in the final
state where $j=u(\bar{u)},d(\bar{d}),c(\bar{c}),s(\bar{s}),b(\bar{b})$
quarks within the SM. We calculated the total background cross section
to be around $4.5\times10^{-5\:}pb$ for $\sqrt{s}\approx3.5\:TeV$
and $12.5\times10^{-5}\:pb$ for $\sqrt{s}\approx5\:TeV$ energy options.
Dominant background contribution in SM is obtained from the tree level
multi-jets + lepton productions where we have \textit{4 b-jets} tagged
in the final state. Second main contribution is obtained by $t\bar{t}$
+ 1jet + 1 lepton where QCD interactions play important role as well.
Inclusively produced two top quarks decay to W\textpm b(bbar) and
W bosons decay hadronically giving at least 2 b-jets, one can obtain
\ensuremath{\ge} 4 b-jets in the final state. Similarly, we have added
the contributions of $t(\bar{t})W^{-(+)}$ + 1jet + 1 lepton and $W^{+(-)}W^{-(+)}$
+ 1jet + 1 lepton processes in the same investigation and entitled
all of these as ``All Top \& W Inclusive'' in Table 3. Third contribution
is obtained by electroweak neutral productions such as ZZ / ZH + 1jet
+ 1 lepton as shown in Table 3. Due to the basic transverse momentum
cut applied at 20 GeV for low-pt jets, gluon jets and a small portion
of quark jets have been removed. Both signal and background events
are produced by setting the factorization and renormalization scales
at 125 GeV with standard NN23LO1 parton distribution function set.
In productions, b-tagging efficiency is assumed to be \%60 and considered
1\% of light-jets faking the leptons while for c-quark jets, the same
fake rate is \%10.

\begin{figure}
\includegraphics[scale=0.28]{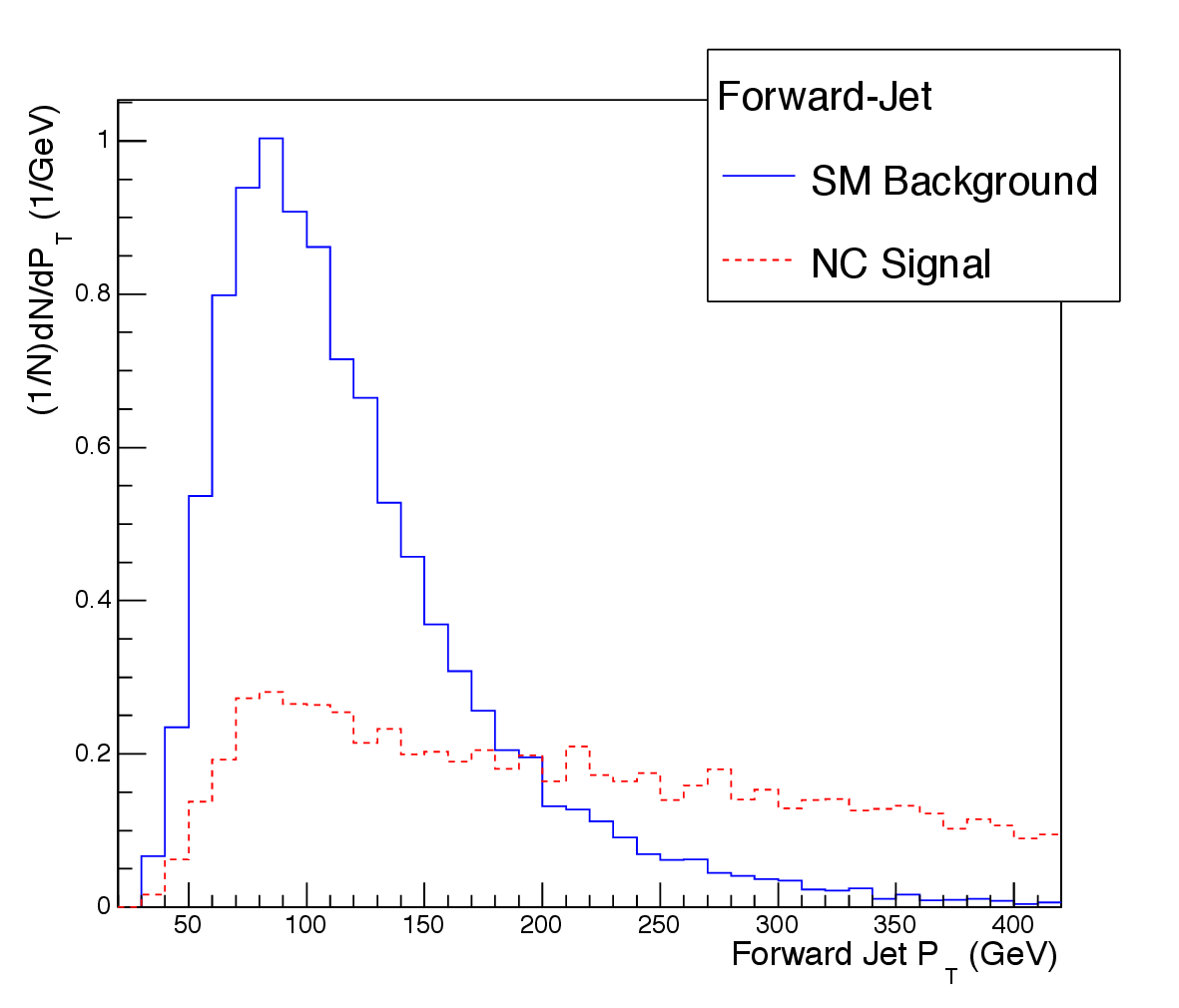}\includegraphics[scale=0.28]{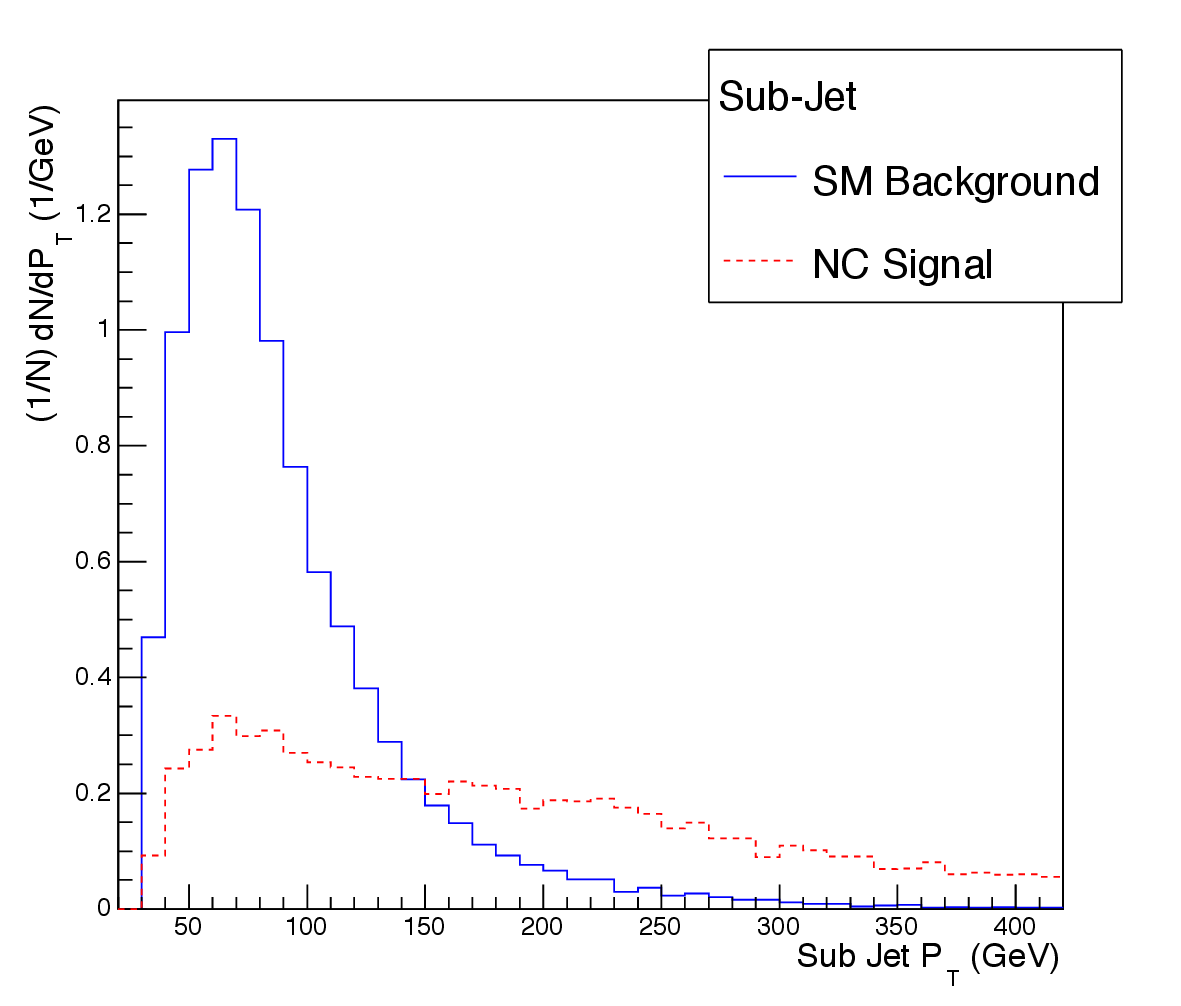}\includegraphics[scale=0.28]{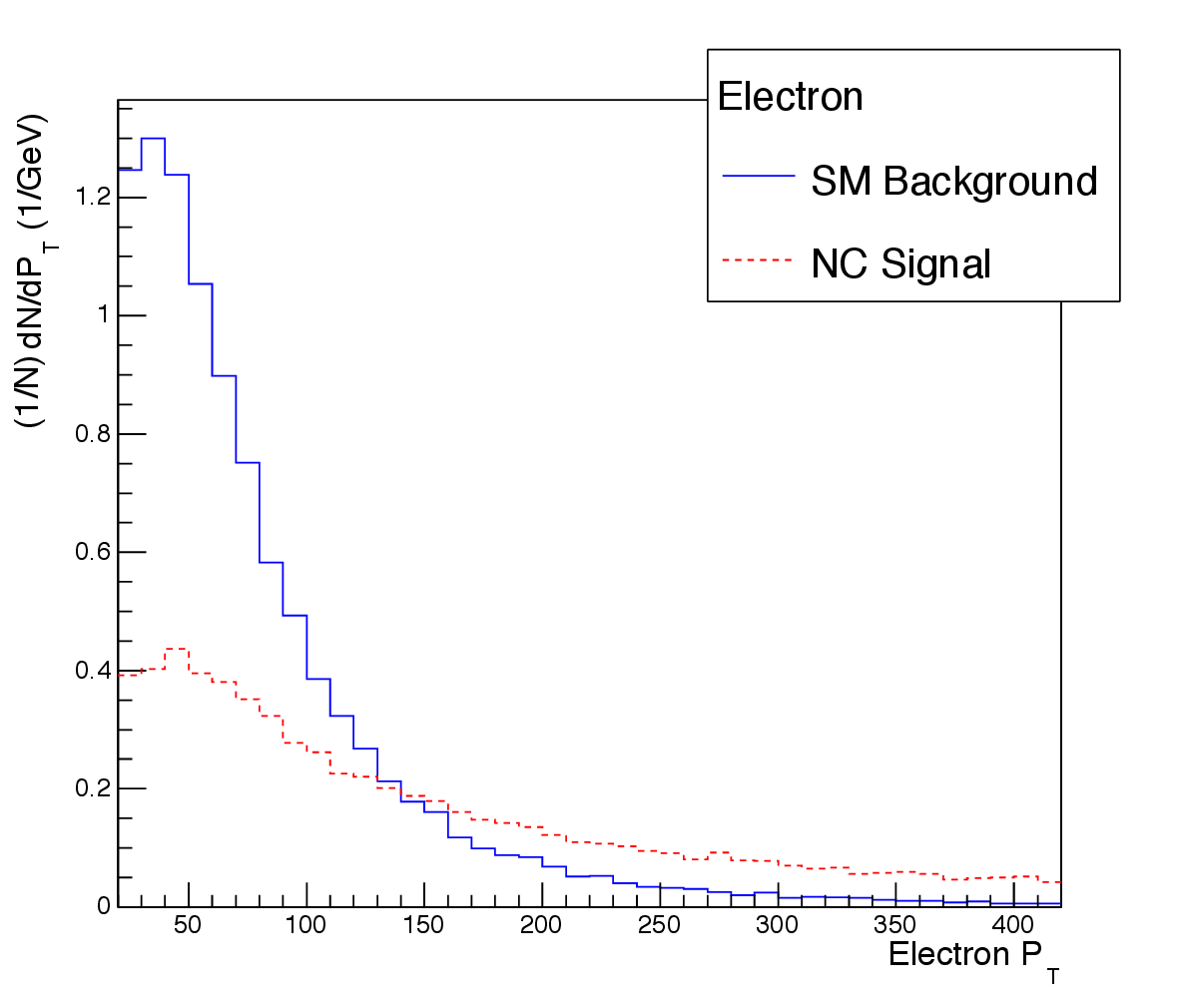}

\caption{Transverse momentum of forward and sub jets for SM background (blue)
and signal (red) respectively.}
\end{figure}

\begin{figure}
\includegraphics[scale=0.28]{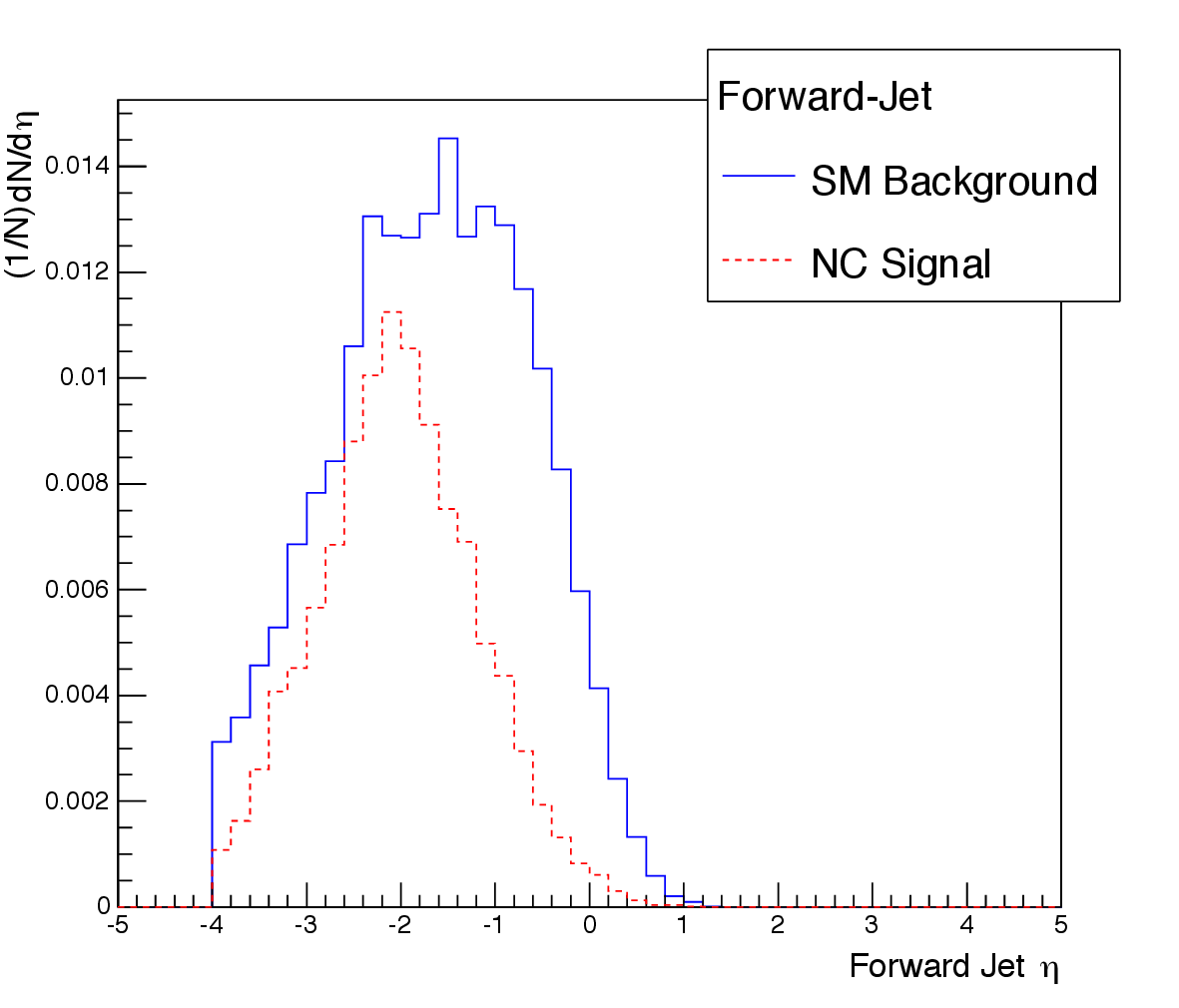}\includegraphics[scale=0.28]{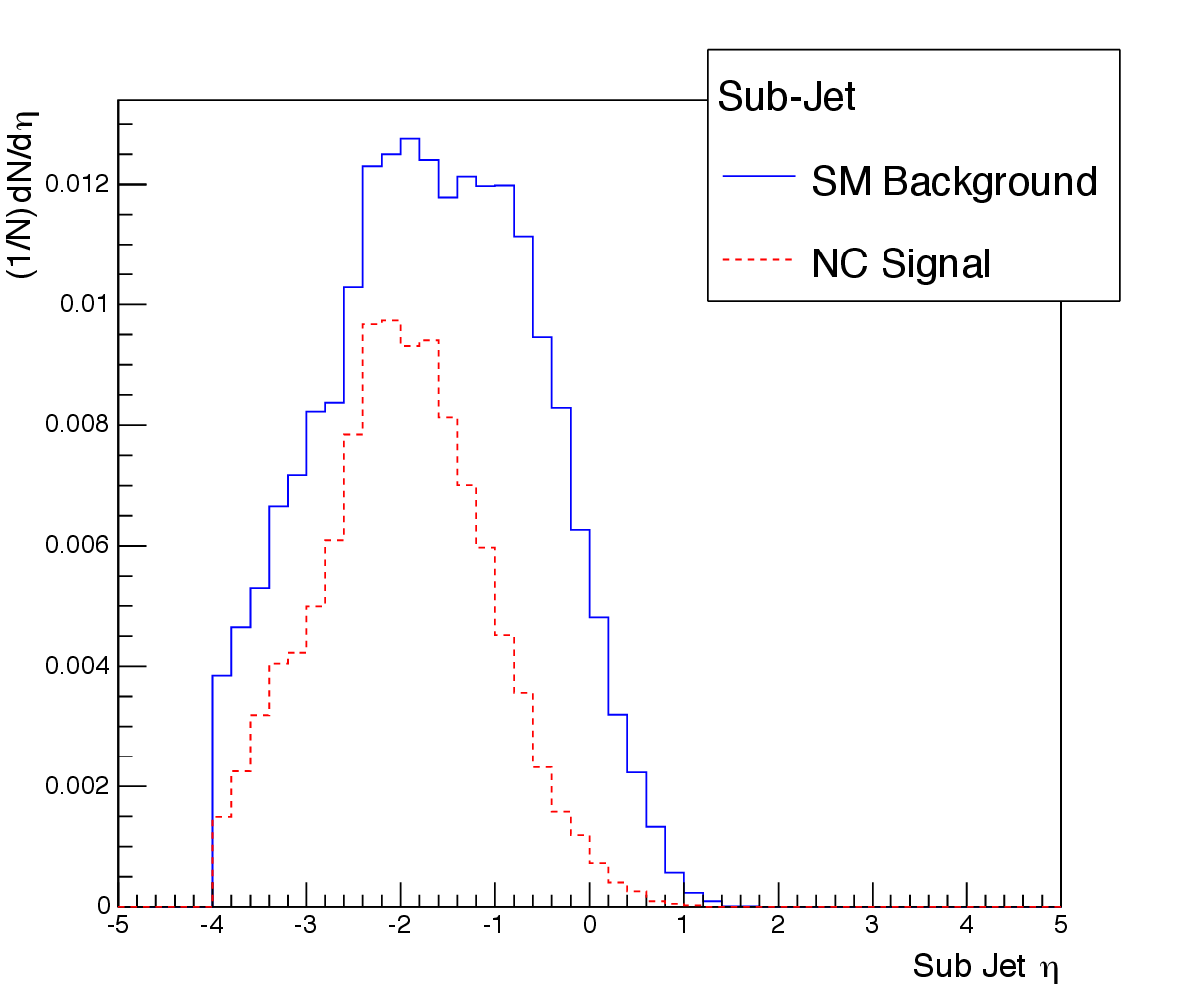}\includegraphics[scale=0.28]{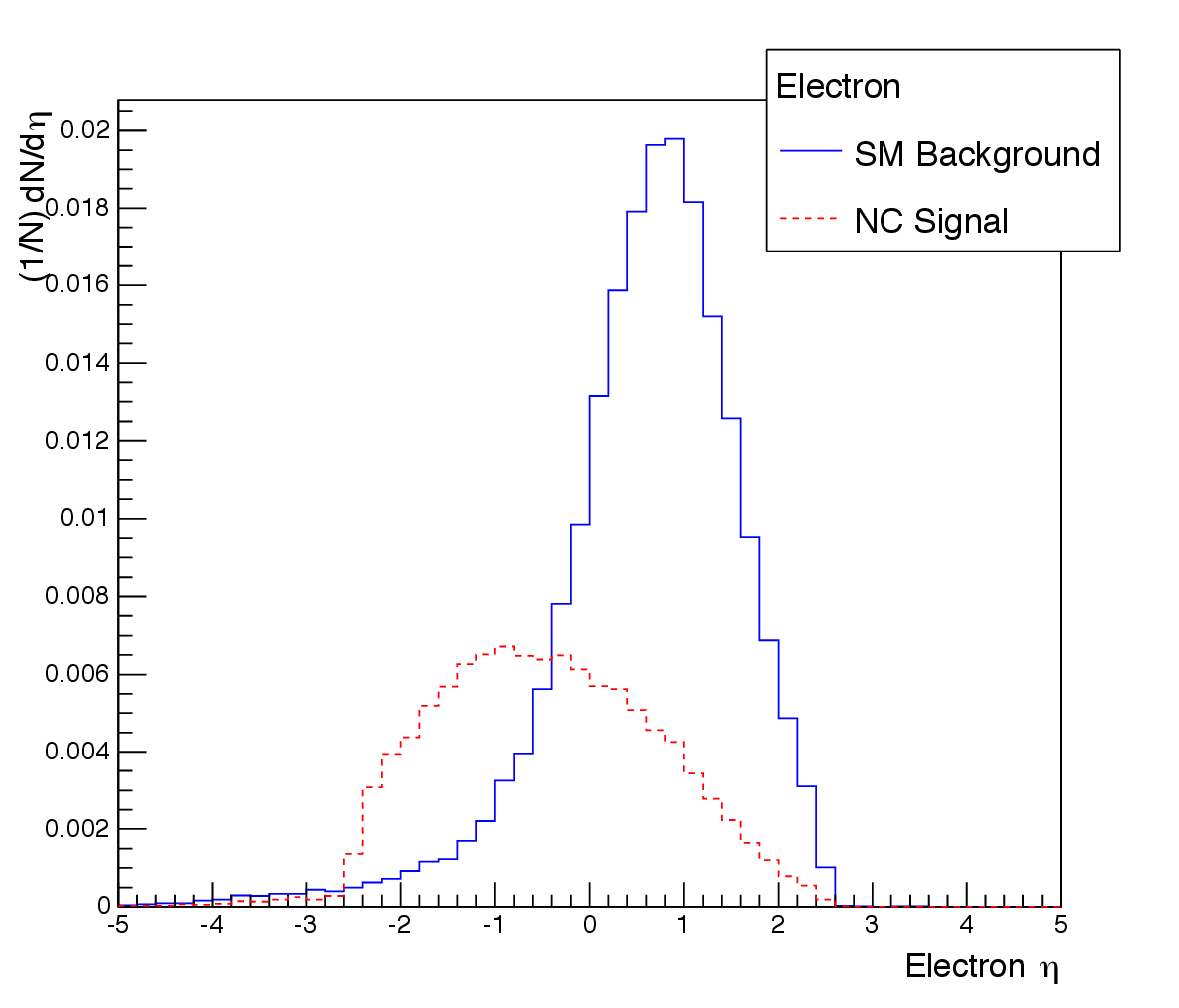}

\caption{Pseudo-rapidity distributions of leading jet, sub-leading jet and
electron for SM background (blue) and signal (red) respectively. }
\end{figure}

\begin{figure}
\includegraphics[scale=0.8]{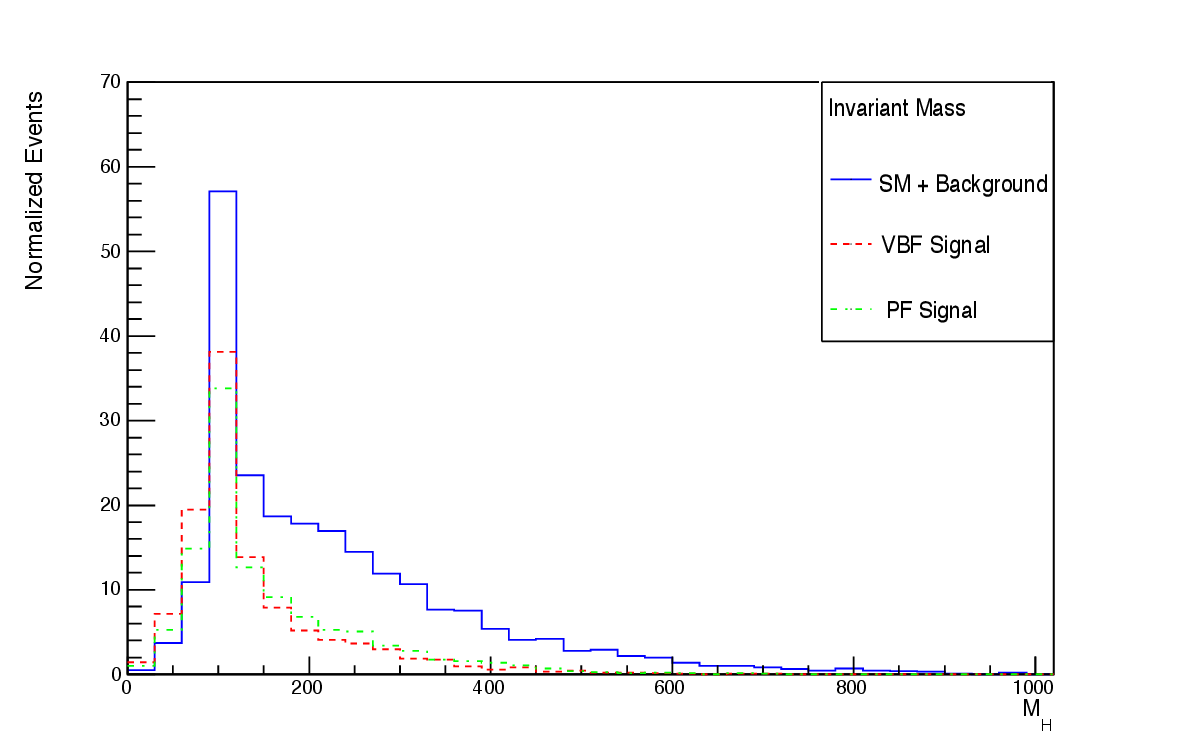}

\caption{Reconstructed invariant mass distribution of Higgs from two b-jets
for SM background (blue), VBF signal (red) and PF signal (green). }
\end{figure}

\begin{figure}
\includegraphics[scale=0.85]{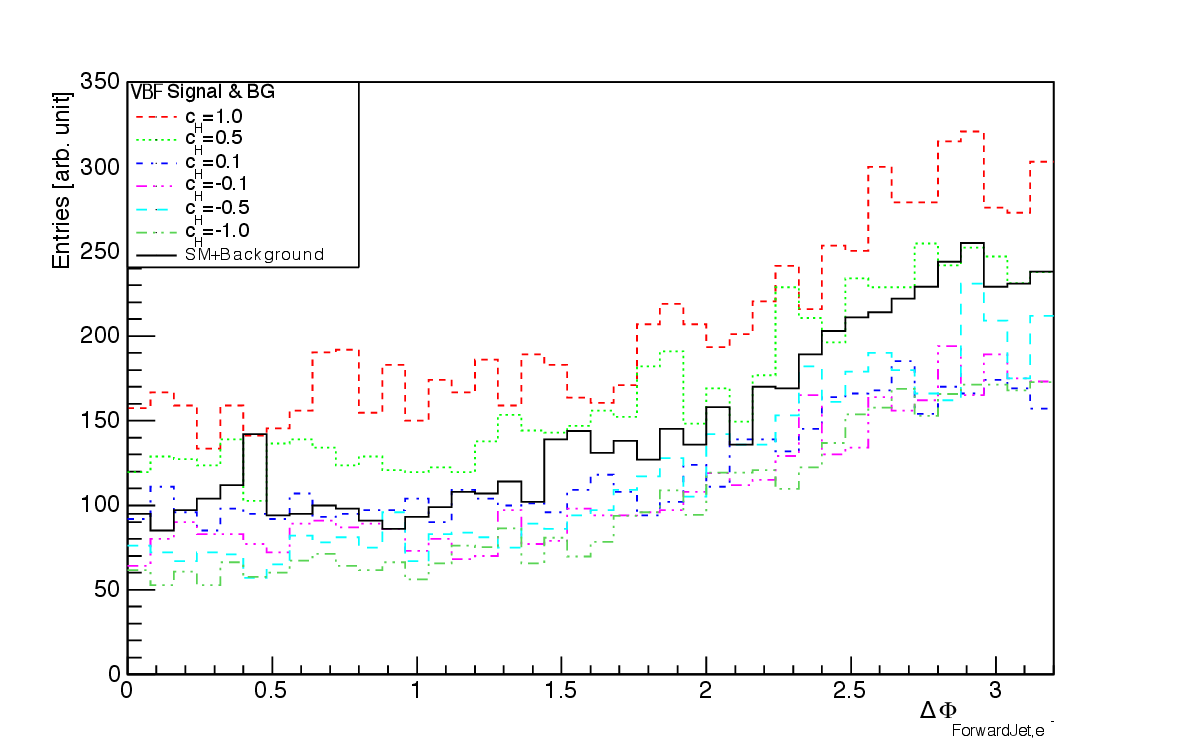}

\caption{Azimuthal angle distribution of lepton and forward jet of VBF signal
for $\bar{c}_{H}=\pm0.1,\pm0.5,\pm1$ with SM + background (black)
for $\sqrt{s}\approx5\:TeV$ option.}
\end{figure}

\begin{figure}
\includegraphics[scale=0.85]{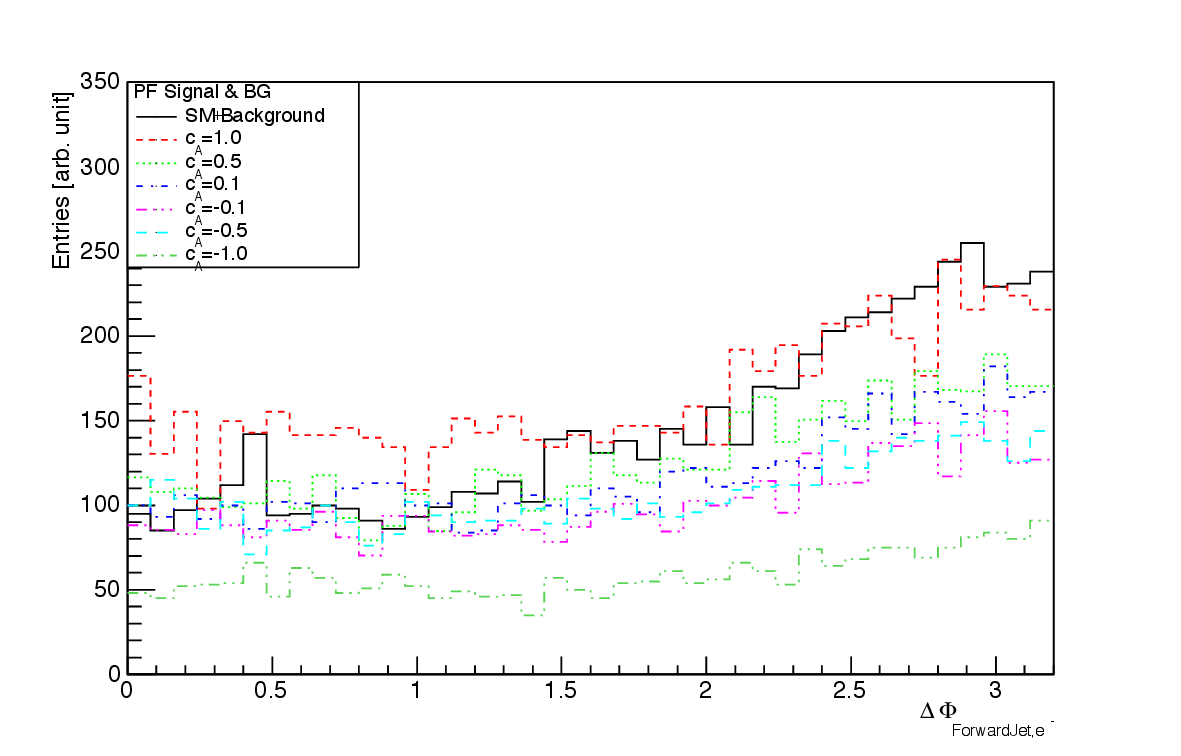}

\caption{Azimuthal angle distribution of lepton and forward jet of PF signal
for $\bar{c}_{\gamma}=\pm0.1,\pm0.5,\pm1$ with SM + background (black)
for $\sqrt{s}\approx5\:TeV$ option.}
\end{figure}

\begin{table}
\caption{Signal and Background events for $\mathcal{L}_{int}=10\:ab^{-1}$ }

\begin{tabular*}{16.5cm}{@{\extracolsep{\fill}}cccccc}
\toprule 
\multirow{2}{*}{} & \multirow{2}{*}{{\footnotesize{}$\sqrt{s}\approx3.5\:TeV$ }} & \multirow{2}{*}{{\footnotesize{}$\sqrt{s}\approx5\:TeV$ }} & \multirow{2}{*}{} & \multirow{2}{*}{{\footnotesize{}$\sqrt{s}\approx3.5\:TeV$(with cuts) }} & \multirow{2}{*}{{\footnotesize{}$\sqrt{s}\approx5\:TeV$ (with cuts)}}\tabularnewline
 &  &  &  &  & \tabularnewline
\midrule
\midrule 
VBF Signal: %
\begin{tabular}{c}
{\scriptsize{}$c_{H}=1.0$}\tabularnewline
{\scriptsize{}$c_{H}=0.5$}\tabularnewline
{\scriptsize{}$c_{H}=0.1$}\tabularnewline
\end{tabular} & {\footnotesize{}}%
\begin{tabular}{c}
{\footnotesize{}1333}\tabularnewline
{\footnotesize{}493}\tabularnewline
{\footnotesize{}169}\tabularnewline
\end{tabular} & {\footnotesize{}}%
\begin{tabular}{c}
{\footnotesize{}3493}\tabularnewline
{\footnotesize{}1383}\tabularnewline
{\footnotesize{}561}\tabularnewline
\end{tabular} &  & {\footnotesize{}}%
\begin{tabular}{c}
{\footnotesize{}224}\tabularnewline
{\footnotesize{}102}\tabularnewline
{\footnotesize{}55}\tabularnewline
\end{tabular} & {\footnotesize{}}%
\begin{tabular}{c}
{\footnotesize{}668}\tabularnewline
{\footnotesize{}352}\tabularnewline
{\footnotesize{}213}\tabularnewline
\end{tabular}\tabularnewline
\midrule
\midrule 
PF Signal: %
\begin{tabular}{c}
{\scriptsize{}$c_{\gamma}=1.0$}\tabularnewline
{\scriptsize{}$c_{\gamma}=0.5$}\tabularnewline
{\scriptsize{}$c_{\gamma}=0.1$}\tabularnewline
\end{tabular} & {\footnotesize{}}%
\begin{tabular}{c}
{\footnotesize{}876803}\tabularnewline
{\footnotesize{}22471}\tabularnewline
{\footnotesize{}8651}\tabularnewline
\end{tabular} & {\footnotesize{}}%
\begin{tabular}{c}
{\footnotesize{}2571242}\tabularnewline
{\footnotesize{}712902}\tabularnewline
{\footnotesize{}26921}\tabularnewline
\end{tabular} &  & {\footnotesize{}}%
\begin{tabular}{c}
{\footnotesize{}428204}\tabularnewline
{\footnotesize{}10544}\tabularnewline
{\footnotesize{}4353}\tabularnewline
\end{tabular} & {\footnotesize{}}%
\begin{tabular}{c}
{\footnotesize{}1496202}\tabularnewline
{\footnotesize{}384701}\tabularnewline
{\footnotesize{}15764}\tabularnewline
\end{tabular}\tabularnewline
\midrule
\midrule 
{[}Backgrounds{]} &  &  &  &  & \tabularnewline
\midrule
\midrule 
{\footnotesize{}4 bjets + 1 jet + 1 e-} & {\footnotesize{}120343} & {\footnotesize{}258911} &  & {\footnotesize{}2} & {\footnotesize{}11}\tabularnewline
\midrule
\midrule 
{\footnotesize{}All Top \& W Inclusive} & {\footnotesize{}82787} & {\footnotesize{}216209} &  & {\footnotesize{}349} & {\footnotesize{}975}\tabularnewline
\midrule
\midrule 
{\footnotesize{}Z / H + 2bjets + 1 jet + 1 e-} & {\footnotesize{}1634.2} & {\footnotesize{}38264} &  & {\footnotesize{}22} & {\footnotesize{}89}\tabularnewline
\midrule
\midrule 
{\footnotesize{}Z / ZZ + 1 jet + 1 e- } & {\footnotesize{}1625.4} & {\footnotesize{}2760} &  & {\footnotesize{}55} & {\footnotesize{}116}\tabularnewline
\midrule
\midrule 
{\footnotesize{}ZH + 1 jet + 1 e-} & {\footnotesize{}407} & {\footnotesize{}690} &  & {\footnotesize{}24} & {\footnotesize{}58}\tabularnewline
\midrule
\midrule 
Total Background & {\footnotesize{}206796.6} & {\footnotesize{}516834} &  & {\footnotesize{}452} & {\footnotesize{}1249}\tabularnewline
\midrule
\midrule 
$\frac{S}{\sqrt{B}}$ for VBF &  &  &  & {\footnotesize{}}%
\begin{tabular}{c}
{\footnotesize{}10.5}\tabularnewline
{\footnotesize{}4.8}\tabularnewline
{\footnotesize{}2.58}\tabularnewline
\end{tabular} & {\footnotesize{}}%
\begin{tabular}{c}
{\footnotesize{}18.9}\tabularnewline
{\footnotesize{}9.96}\tabularnewline
{\footnotesize{}6.03}\tabularnewline
\end{tabular}\tabularnewline
\midrule
\midrule 
$\frac{S}{\sqrt{B}}$ for PF &  &  &  & {\footnotesize{}}%
\begin{tabular}{c}
{\footnotesize{}$2\times10^{4}$}\tabularnewline
{\footnotesize{}496}\tabularnewline
{\footnotesize{}204}\tabularnewline
\end{tabular} & {\footnotesize{}}%
\begin{tabular}{c}
{\footnotesize{}$4.2\times10^{4}$}\tabularnewline
{\footnotesize{}$10^{4}$}\tabularnewline
{\footnotesize{}446}\tabularnewline
\end{tabular}\tabularnewline
\bottomrule
\end{tabular*}
\end{table}

\section{EVENT SELECTION AND ANALYSIS}

Event selection criteria: (1) Four b-tagged jets and a light jet is
selected with $p_{T}>20\:GeV$. (2) $|\eta|<5$ for all jets and $|\eta|<2.5$
for leptons applied. (3) Between jets and b-jet and a lepton $\Delta R=0.4$
applied. (4) Event selection cut: $p_{T}>150\:GeV$ for leading jet
$p_{T}>110\:GeV$ for sub jets. (5) Invariant mass window cut for
both b-jet pairs: $m_{bb}\:\epsilon\:[50,130]$. (6) Vetoing events
if missing transverse energy, $E_{T}$ $>20\:GeV$.

In Fig.4 and 5, we present the kinematic distributions in comparison
with background and SM processes through a Z boson mediator while
$c_{H}=0.1$ . Fig. 4 shows that the forward and sub jets in signal
have a separable transverse momentum than the background jets as expected,
while the pseudo-rapidity distributions (Fig.5) behave similar for
both signal and background jets. For outgoing electron, one can see
that the signal distribution slightly deviates to the negative region,
while background signal locates at zero pseudo-rapidity. For the higher
values of $\bar{c_{H}}$ (or $\bar{c_{\gamma}}$ in PF process), the
higher negative deviation for the mean of pseudo-rapidity distributions
are observed. In Fig.6, it is seen the reconstructed invariant mass
distribution of one Higgs boson ($M_{H})$ from two b-jets within
di-Higgs for both signal and background located at around 125 GeV.
This also shows that in SM, the large contribution comes from the
b-jets created by di-Higgs decays. As a significant evidence for coupling
seperations, azimuthal angle distribution ($\Delta\phi$) between
lepton and forward jet for VBF and PF signals are shown in Fig.7 and
Fig.8, respectively for $\sqrt{s}\approx5\:TeV$. Similarly for $\sqrt{s}\approx3.5\:TeV$,
same distributions are obversed with a factor $\sim0.286$ in the
event count. About the shape of the distribution of $\Delta\phi$,
we observed that this interference is strictly dependent on the coefficients
and detector parameters. For evaluating the limits in Fig.9, we calculated
statistical significances and followed the methods described in Ref.\citep{Statistics}
It's also worthwile to comment on the event selection criteria: First
three conditions consist of almost default cuts for event production.
Fourth condition has higher $p_{T}$ cuts that can be seen directly
from Fig.4 to separate signal from background. Fifth condition is
extremely effective if b-jets are not produced by a higgs decay. Sixth
condition is also an important selection criteria especially for removing
top background since W+/W- bosons from a top quark decay emerge neutrinos
while decaying to leptons. For hadronic decay of W bosons, note that
$Br(W^{-}/W^{+}\rightarrow\bar{c}b/c\bar{b})\backsim0.01$ leaves
fairly less b-pairs in the final state. We have determined cuts after
obtained kinematic distributions as Fig.4-6 and optimised scanning
over variables to obtain the highest significance, namely $\frac{S}{\sqrt{B}}$
where S signal events and B background events. The last two columns
of Table 3, denote the survived event numbers after applying our selection
criteria. 

To evaluate above predictions within the realistic perspective, one
should consider reconstruction efficiencies from the detector, systematic
errors on luminosity measurements and pile-up treating. Note that
along with the b-tagging efficiency, W/Z reconstruction efficiencies
and uncertainities in decay channels may affect sentivity results
as mentioned for LHC in Ref \citep{key-30}. These uncertainities
which are obtained directly from previous experiments, will likely
to improve with integrated luminosity. For ease of comparison, while
a reduction of total statistical and experimental systematic uncertainties
by a factor of about 0.3 for $L_{int}=300\:fb^{-1}$ and about 0.1
for $L_{int}=3000\:fb^{-1}$ for LHC, one should expect a lower factor
extrapolating the same idea for FCC-he. 

\section{CONCLUSION}

In this letter, we have investigated the sensitivity on the Higgs
boson couplings and Wilson coefficients in productions through NC
mechanism (in Fig. 1) for FCC-he. Since the process is possible through
both $Z$ boson and $\gamma$ mediators, one should take into account
the interference of VBF and PF processes with the right parameter
set. We observed that $\Delta\phi$ variable which is strictly affected
by detector parameters, is a key to separate interferences of both
VBF and PF signals. It is observed that di-Higgs production through
NC mechanism has a major sensitivity to $\bar{c}_{\gamma}$ and $\bar{c}_{H}$
coefficients within the considerations of electroweak precision measurements.
FCC-he collider can cover $\bar{c}_{H}\:(\bar{c}_{\gamma})$ coefficients
as in Fig.9 through NC processes with integrated limunosities up to
$3\:(50)\:ab^{-1}$ respectively. On the other hand, one can reveal
the corresponding Higgs couplings by obtaining limits of Wilson parametrization
that is involved in Higgs productions at a specific limunosity. Thus,
we present $g_{hh\gamma\gamma}/\tilde{g}_{hh\gamma\gamma}$, $g_{hhzz}$
and $\tilde{g}_{hhzz}$ in Fig. 10 that shows the required limunosities
to discover these couplings. An integrated luminosity of $10\:ab^{-1}$
can set limits on these couplings {[}-0.0005, 0.0005{]}, {[}-0.05,
0.05{]} and {[}-0.012, 0.012{]} respectively. Although our approach
is based on the single parameter dominance hypothesis, considering
the Wilson coefficients, one can compare the $\bar{c}_{H}\:(\bar{c}_{\gamma})$
constraints obtained by LHC data in Ref \citep{key-30}. It is seen
that LHC (with $L_{int}=3000\:fb^{-1}$) targeted to set similar limits
on compared coefficients $\bar{c}_{H}\:(\bar{c}_{\gamma})$ with the
FCC-he collider at the same luminosity levels. However, this comparison
of different types of colliders should be made in the context of uncertainities
and systematic errors on which FCC-he may have some advantages as
a future collider.

\begin{figure}
\includegraphics[scale=0.65]{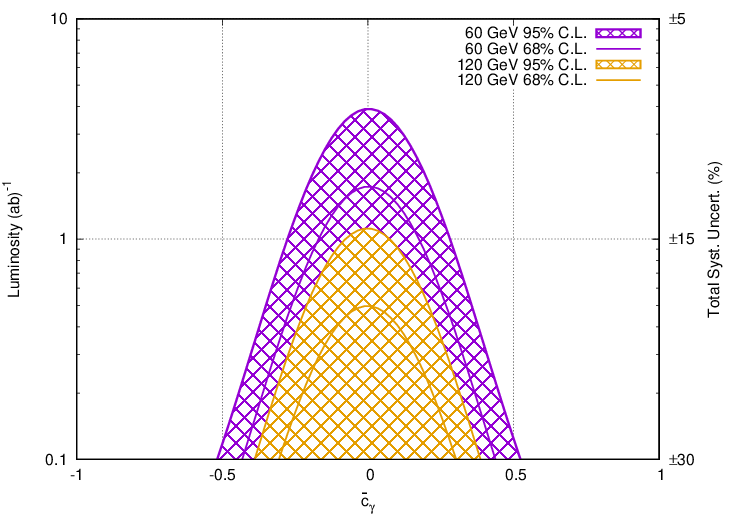}\includegraphics[scale=0.65]{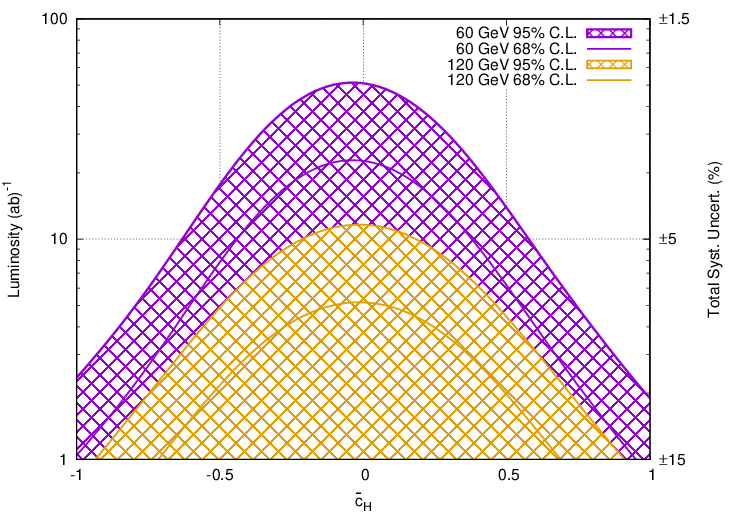}

\caption{Required integrated luminosities to obtain the limits on Wilson coefficients
$\bar{c}_{H},\:\bar{c}_{\gamma}$ for 60 GeV and 120 GeV electron
beam energy options at FCC-he where the shaded areas are not allowed
assuming $\bar{c}_{T},\bar{c}_{W},\bar{c}_{B}=10^{-3}$ and all other
Wilson coefficients are zero. Total systematic uncertainties are roughly
extrapolated from LHC data in percentage. }
\end{figure}

For EFT approach, it is known that above the new physics scale, $\varLambda$,
Lagrangian expansions will be unconvinced and limits on the couplings
will deteriorate rapidly. One can see from the Wilson parametrization
that the coefficients can naively be expressed in terms of M, overall
mass scale. However, couplings such as $\tilde{g}_{hh\gamma\gamma}$,
$\tilde{g}_{hhzz}$ that have degrading sensitivities because of the
higher order dependences, have deterioration of limits such that,
at higher energies, deviations from the original limit in percentage
getting lower. Thus, according to recent mass limits on heavy particles,
one can see that the deterioration of limits cannot deviate above
\%10. Although similar detailed searches at the LHC are avaliable
to set limits on corresponding Higgs couplings, it is possible to
obtain high precision on the couplings using FCC-he advantages in
center of mass energy and background. 

\section*{ACKNOWLEDGEMENT}

We would like to thank especially Dr. Uta Klein, Dr. Bruce Mellado,
Dr. Rob Apsimon and Dr. Orhan Cakir for their comments, fruitful discussions
for this work. 

\begin{figure}
\subfloat[]{\includegraphics[scale=0.75]{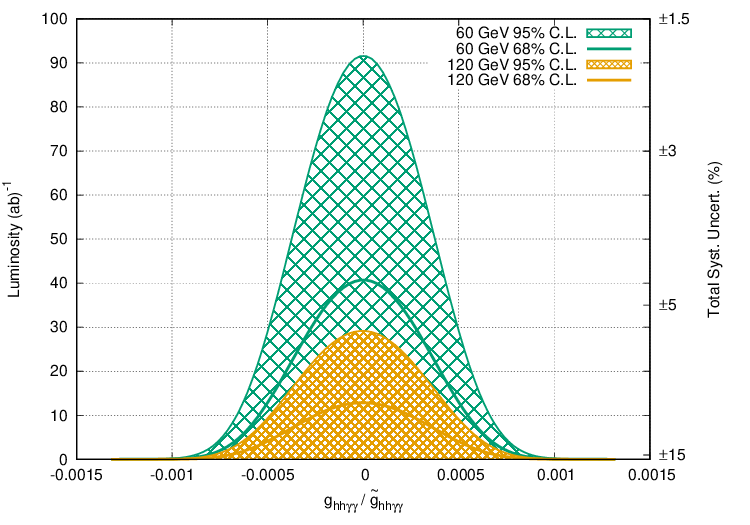}

}\hspace{1\paperwidth}\subfloat[]{\includegraphics[scale=0.65]{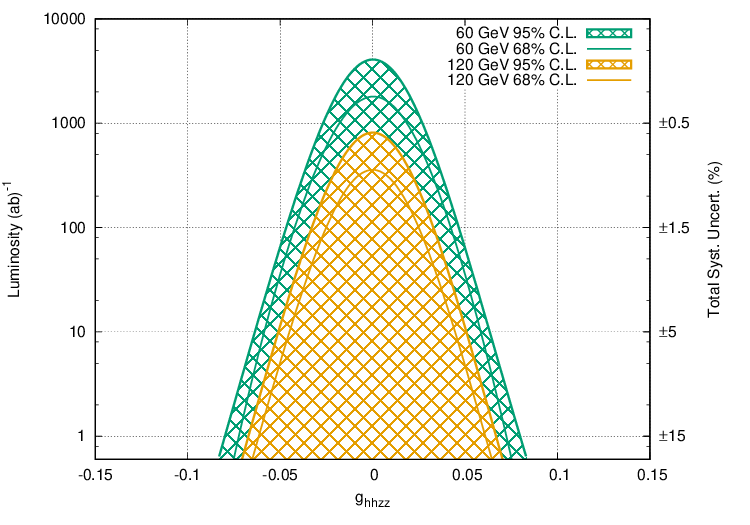}\includegraphics[scale=0.65]{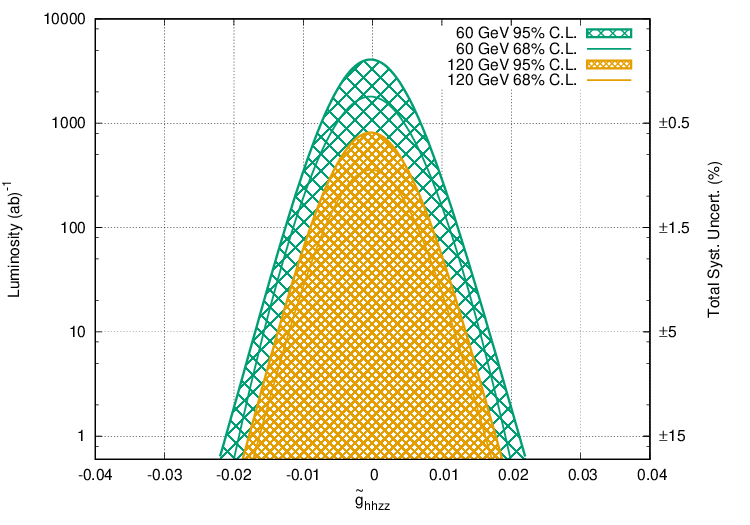}

}

\caption{Required integrated luminosities to obtain the limits on the corresdonding
couplings (a) $g_{hh\gamma\gamma}/\tilde{g}_{hh\gamma\gamma}$ (b)
$g_{hhzz}$ and $\tilde{g}_{hhzz}$ for 60 GeV and 120 GeV electron
beam energy options at FCC-he where the shaded areas are not allowed
assuming $\bar{c}_{T},\bar{c}_{W},\bar{c}_{B}=10^{-3}$ and all other
Wilson coefficients are zero. Total systematic uncertainties are roughly
extrapolated from LHC data in percentage. }
\end{figure}

\end{document}